\documentclass[prd,showpacs,amsmath,showkeys,twocolumn,floatfix,amssymb, preprintnumbers, nofootinbib, superscriptaddress]{revtex4} 
\usepackage{epsfig,dcolumn}
\usepackage{graphicx}
\usepackage{comment} 
\usepackage[usenames]{color}
\usepackage{graphicx}
\usepackage{bm}
 \usepackage{ifpdf}

\begin{document}

\title{ Multiple-particle interaction  in $1+1$ dimensional lattice model }

\author{Peng~Guo}
\email{pguo@csub.edu}

\affiliation{Department of Physics and Engineering,  California State University, Bakersfield, CA 93311, USA}
\affiliation{Kavli Institute for Theoretical Physics, University of California, Santa Barbara, CA 93106, USA}

\author{Tyler~Morris}

\affiliation{Department of Physics and Engineering,  California State University, Bakersfield, CA 93311, USA}

\date{\today}

\begin{abstract} 
  Finite volume multiple-particle interaction is studied in a  two-dimensional complex $\phi^4$    lattice model. The existence of  analytical solutions to the $\phi^4$ model in two-dimensional space and time   makes it a perfect model for the numerical study of finite volume effects of multi-particle interaction.    The spectra from multiple particles  are extracted from the Monte Carlo simulation on various lattices in several moving frames. The  $S$-matrix  of multi-particle scattering in $\phi^4$ theory is completely determined  by two fundamental parameters: single particle mass and the coupling strength of two-to-two particle interaction. These two parameters are fixed by studying single-particle and two-particle spectra. Due to the absence of  the diffraction effect in the $\phi^{4}$ model, three-particle quantization conditions are given in  a simple analytical form. The three-particle spectra from simulation  show remarkable agreement with the  prediction of exact solutions. 
  \end{abstract} 


\maketitle

\section{Introduction}
\label{intro} 
One of the outstanding but challenging goals in nuclear/hadron physics is to understand the dynamics of particle interaction. Multiple particle interaction is not only important to nuclear/hadron physics, but also plays a crucial role in astrophysics, atomic and condensed matter physics. However, the complication increases dramatically with increasing numbers of dynamical degrees of freedom and poses a significant obstacle in studying and understanding multi-particle interaction. Fortunately, the simplest case of multi-particle interaction turns out to be manageable, three-particle interaction.  The dynamics of three-particle interaction were well developed and  studied in the past  \cite{Taylor:1966zza,Basdevant:1966zzb,Gross:1982ny,Faddeev:1960su,Faddeev:1965,Phillips:1966zza,Gloeckle:1983,Fedorov:1993,Gloeckle:1995jg,Khuri:1960zz,Bronzan:1963xn,Aitchison:1965kt,Aitchison:1965zz,Aitchison:1966kt,Pasquier:1968zz,Pasquier:1969dt,Guo:2014vya,Guo:2014mpp,Danilkin:2014cra,Guo:2015kla}. Recent progress in high statistic experiments, such as GlueX and CLAS programs, have triggered   renewed interests in three-body dynamics. One example is  the extraction of $u$- and $d$-quark mass difference from  \mbox{$\eta \rightarrow 3 \pi$}  decay process \cite{Kambor:1995yc,Anisovich:1996tx,Colangelo:2009db,Lanz:2013ku,Schneider:2010hs,Kampf:2011wr,Guo:2015zqa,Guo:2016wsi}.  On the other hand, lattice QCD provides an unprecedented opportunity for the study of multiple particle interaction from the heart of hadrons with quarks and gluons as the fundamental building blocks. Recent advances in lattice computation have made the study of hadron interaction especially possible \cite{Aoki:2007rd,Sasaki:2008sv,Feng:2010es,Dudek:2010ew,Beane:2011sc,Lang:2011mn,Aoki:2011yj,Dudek:2012gj,Dudek:2012xn,Wilson:2014cna,Wilson:2015dqa,Dudek:2016cru}. Because lattice QCD is formulated in Euclidean space, access to scattering information is not always direct. That adds some additional complication in multi-particle studies in lattice QCD  as well as the intense numerical computation and other difficulties. A formalism was proposed nearly 30 years ago by L\"uscher   \cite{Lusher:1991} to tackle the two-particle elastic scattering problem in finite volume; it is known as the L\"uscher  formula. Since then,   the framework quickly extended to   moving frames  \cite{Gottlieb:1995,Lin:2001,Christ:2005,Bernard:2007,Bernard:2008},  and to     coupled-channel scattering \cite{Liu:2005,Lage:2009,Doring:2011,Aoki:2011,Briceno:2012yi, Hansen:2012tf,Guo:2013cp}. In the three-particle sector, many groups have made remarkable progress  \cite{Kreuzer:2008bi,Kreuzer:2009jp,Kreuzer:2012sr,Polejaeva:2012ut,Briceno:2012rv,Hansen:2014eka,Hansen:2015zga,Hansen:2016fzj,Hammer:2017uqm,Hammer:2017kms,Guo:2016fgl,Guo:2017ism,Meissner:2014dea,Briceno:2017tce,Sharpe:2017jej,Mai:2017bge,Guo:2017crd} related to the theoretical algorithm of extracting scattering amplitudes from lattice data in recent years.  

A three-particle lattice simulation was recently performed  based on a complex $\phi^{4}$ toy model \cite{Romero-Lopez:2018rcb}, the data analysis was carried out by adopting  effective theory framework. However, the simulation and analysis are limited solely to ground state energy levels where all three particles are nearly at rest, and the three-particle signals are quite noisy. In present work, we aim to perform a simulation on  multiple-particle interaction also using $\phi^{4}$ model,   and  study the finite volume effect on multiple-particle spectra in a better controlled environment and a more systematic way.  For this purpose, multiple numbers of multi-particle operators are used in our simulation and variational analysis \cite{Michael:1985ne,Luscher:1990ck,Blossier:2009kd} is implemented to extract   excited state energy levels.  The exact scattering solutions of $\phi^{4}$ theory in $1+1$ dimensions are known in both free space   \cite{Thacker:1974kv,McGuire:1964zt,Yang:1967bm} and finite volume \cite{Guo:2016fgl}. Taking advantage of existing analytic multiple-particle scattering solutions,    the simulation is therefore performed in $1+1$ dimensional space and time for various lattice sizes and moving frames. The exact scattering solutions are used in   data analysis  of multi-particle simulation. In principle, the multiple particle scattering $S$-matrices are completely determined by only two free parameters: single particle mass and coupling strength of two-to-two particle interaction. The single particle mass is obtained from single particle correlation functions, and the coupling strength of pair-wise interaction is extracted by studying two-particle scattering spectra in a lattice. The comparison between three-particle scattering spectra and predicted three-particle energies by using analytic expression of three-particle quantization conditions are presented in the end.

The paper is organized as follows. The exact solutions of $\phi^4$ theory for two-body and three-body interaction are  summarized    in Section \ref{phi4sols}.   The algorithm of the Hybrid Monte Carlo simulation  of lattice model and strategy of data analysis are briefly discussed in Section \ref{lattmodel}. The construction of multi-particle operators,    multi-particle spectra in lattice simulation and   data analysis   are described in Section  \ref{spectra}. The summary and outlook are given in Section \ref{summary}.

\section{Exact solution of $\phi^4$ model in $2D$}\label{phi4sols}
In this section, we summarize some    results of the two-dimensional $\phi^4$ model. Classical action of the complex $\phi^4$ model in two-dimensional  Euclidean space   is 
\begin{equation}
S=   \int d^{2}x \left [\frac{1}{2}    \partial \phi^*   \partial \phi   + \frac{1}{2} \mu^{2} | \phi|^{2} + \frac{g }{4!} |\phi|^{4} \right ]  , \label{phi4action}
 \end{equation}
where $x=(x_{0},x_{1})$ are temporal and spatial coordinates in two-dimensional Euclidean space,  respectively. It is known \cite{Thacker:1974kv} that  the complex $\phi^4$ model in Eq.(\ref{phi4action}) is equivalent to  a non-relativistic one-dimensional $N$-body  interaction problem of particles interacting with pair-wise $\delta$-function   potentials, 
\begin{align}
H =  - \frac{1}{2 m} \sum_{i=1}^N \frac{\partial^2}{\partial x_{1,i}^2} + V_0  \sum_{i<j} \delta(x_{1,i} - x_{1,j}),
\end{align}
where $x_{1,i}$ refers to the spatial position of $i$-th particle, and $m$ stands for the mass of identical bosons. The coupling strength of $\delta$-function potential, $V_0$, differs from renormalized $g$ in Eq.(\ref{phi4action}) by a constant factor.
The exact solutions of multi-particle interaction with $\delta$-function potentials were studied and obtained in both free space \cite{Thacker:1974kv,McGuire:1964zt,Yang:1967bm} and finite volume \cite{Guo:2016fgl}. In fact, the particles interacting with $\delta$-function potential in $2D$ is only one of few exactly solvable multi-particle scattering problems.  The multi-particle wave function is described completely by the linear superpositions of plane waves with all possible permutation on particle momenta.  No new momenta are generated by collision, all the diffraction effects are canceled out as the consequence of Bethe's hypothesis \cite{Bethe:1931hc,Lieb:1963rt}. The multi-particle $S$-matrix therefore is factorized into the product of a number of two-particle scattering amplitudes, as if the process of multi-particle scattering would be a succession of separated elastic two-particle collisions \cite{Guo:2016fgl}.

In finite volume for  two-particle scattering, only one quantization condition is required \cite{Guo:2016fgl,Guo:2013vsa}
\begin{equation}
\cot \delta(k) + \cot  \frac{\frac{PL}{2} + k L}{2} =0,
\end{equation}
where $P= p_1 + p_2$ and $k = \frac{p_1 - p_2}{2}$ denote center of mass and relative momenta of two particles. The phase shift $\delta(k)$ for $\delta$-function potential is given by \mbox{$\delta(k) =\cot^{-1} ( - \frac{2 k}{mV_0})$}. The $L$ stands for the size of the square box in $2D$, and center of mass momentum is discretized because of the periodic boundary condition of lattice: $P=\frac{2\pi}{L} d , d \in \mathbb{Z}$.

For three-body scattering in finite volume,  three quantization conditions are obtained \cite{Guo:2016fgl}.  Only two of them are independent,
\begin{align}
&  \cot \left ( - \delta( -  q_{31} )  - \delta ( q_{12}  ) \right )   +  \cot \frac{PL - p_1 L}{2}=0, \nonumber   \\
&  \cot \left (  \delta( -  q_{23} )  + \delta ( q_{12}  ) \right )   +  \cot \frac{PL - p_2 L}{2}=0, \label{3bexactsol}
\end{align}
where  all the relative momenta are given in terms of two independent particle momenta: $p_1$ and $p_2$,  \mbox{$q_{31} = \frac{P- 2 p_1 - p_2  }{2}$}, \mbox{$q_{12} =\frac{p_1-p_2}{2} $}, \mbox{$q_{23} =  \frac{ p_1 + 2 p_2 -P }{2}$}. The momentum of particle-3 is constrained by momentum conservation, \mbox{$p_3= P-p_1-p_2$}. Again, center of mass momentum of three-particle is quantized in the periodic box: $P=\frac{2\pi}{L} d , d \in \mathbb{Z}$.

\section{ The lattice $\phi^4$ model action }\label{lattmodel}
The lattice $\phi^4$ action is obtained from Eq.(\ref{phi4action}) by replacing the continuous derivative with discrete difference:  $\partial  \phi(x) \rightarrow  \phi(x+\hat{n}) - \phi(x)$,  where $\hat{n}$ denotes the unit vector in direction $x_i$ on a  periodic square  lattice. In addition,  by introducing   two new parameters: $\mu^2 = \frac{1-2 \lambda}{\kappa}-8$ and $g=\frac{6 \lambda}{\kappa^2}$, and  also rescaling the $\phi$ field by $\phi \rightarrow \sqrt{2\kappa } \phi$, we thus obtain 
\begin{align}\label{action}
S (\phi)=& -   \kappa  \sum_{x, \hat{n}} \phi^*(x) \phi(x+\hat{n})  + c.c. \nonumber \\
&+ (1-2 \lambda)  \sum_{x} | \phi(x) |^2 + \lambda \sum_{x} | \phi(x) |^4 ,
\end{align}
where $x=(x_{0},x_{1})$ now refers to discrete coordinates of   Euclidean $T\times L$  lattice site.

\subsection{Hybrid Monte Carlo algorithm}
The Hybrid Monte Carlo algorithm \cite{Duane:1987de,Duane:1986iw} is adopted in our numerical simulation, the complex $\phi^4$ model  is treated as a coupled two component scalar field model, $\phi = (\phi_0, \phi_1)$.  In Hybrid Monte Carlo simulation  \cite{Duane:1987de,Duane:1986iw}, an auxiliary Hamiltonian is introduced
\begin{align}
H = \frac{1}{2} \sum_x \pi^*(x) \pi (x) + S (\phi), \label{hamiltonian}
\end{align}
where   $\pi = (\pi_0, \pi_1)$ are fictitious conjugate momenta of $\phi= (\phi_0, \phi_1)$ field.  
The auxiliary Hamiltonian in Eq.(\ref{hamiltonian}) defines classical evolution of both $\pi$ and $\phi$ fields  over a  fictitious time $\tau$ within an interval $  [0, \tau]$:
\begin{align}
&   \phi_i (\tau) =  \phi_i (0) + \int_0^{\tau} d \tau' \pi_i (\tau'),   \nonumber \\
&   \pi_i (\tau) = \pi_i (0)  - \int_0^{\tau} d \tau'  \frac{\partial S(\phi (\tau') )}{ \partial \phi_i (\tau')},  \ \ \ \ i=0,1. \label{eqofmotion}
\end{align}
The trajectory of $(\phi,\pi)$ over   time interval $  [0, \tau]$ is determined by the solutions of motion equations in Eq.(\ref{eqofmotion}).

The two pairs of components, $(\phi_0, \pi_0)$ and $(\phi_1,\pi_1)$, are updated alternately for each sweep over an entire lattice. Updating each pair $(\phi_i, \pi_i )$ is followed with the standard Hybrid Monte Carlo algorithm: 

 (i) the trajectory begins with choosing a random  distribution of  fields $(\phi (\tau) , \pi (\tau) )$ at initial time $\tau=0$. The initial conjugate momenta, $\pi_{i} (0)$, are generated according to the Gaussian probability distribution: \mbox{$P(\pi_i) \propto e^{- \frac{\pi_i^2}{2}}$}. 
 
 (ii)  solve motion equations in Eq.(\ref{eqofmotion}) to evolve $(\phi_i (\tau) , \pi_i (\tau) )$    over the trajectory up to a  time $\tau$.  The motion equations,  Eq.(\ref{eqofmotion}), are solved numerically by the leapfrog method  \cite{Duane:1986iw}.
 
  (iii) accept the proposed new fields, $(\phi (\tau) , \pi (\tau) ) $, with probability:  \mbox{$ P_{acc} = \mbox{Min} \left [ 1 , e^{- \triangle H }\right ]$}, where  \mbox{$ \triangle H = H(\tau) - H(0)$}.

The simulations are performed with the choice of the parameters:  $\kappa =0.1286$, and $\lambda  = 0.01$.  The temporal extent of the lattice is fixed at $T=80$,  and the   spatial extent of lattice, $L$, are from $10$ up to $45$. For each set of lattice size  and moving frame,   one million measurements are generated. The length of trajectory is fixed at $\tau=8$,  the $(\phi,\pi)$ fields evolve from  \mbox{$\tau=0$} up to \mbox{$\tau=8$} over $100$ discrete steps.

\subsection{Strategy of data analysis  }

As already mentioned in Section \ref{phi4sols}, the two-dimensional $\phi^4$ model is exactly solvable, the solutions of the model are given in terms of only two free parameters: particle mass, $m$, and the coupling strength of $\delta$-function potential, $V_0$. The mass of identical particles, $m$, can be extracted from one-particle spectra of the lattice simulation. The second parameter, $V_0$, can be fixed by two-particle spectra from the simulation.   
Taking advantage of  the existence of exact solutions of the two-dimensional  $\phi^4$ model provides an excellent playground and controlled environment for a systematic study of  finite volume effects of  multi-particle scattering  in lattice simulation. In present work, we are not aiming at obtaining any new  fundamental information from three-body spectra, such as three-body force effect, {\it etc.} Instead, after fixing $m$ and $V_0$ from one- and two-particle spectra, we tend to study how well the three-body spectra from simulation match the prediction of exact solutions. In real QCD simulation, the significant difference between simulation results of three-body spectra and prediction based on pair-wise interaction may signal the effect of three-body forces or something more fundamental. The present work serves only as a testbed for more realistic future lattice studies of multi-particle interaction.

To accomplish the goal of this work mentioned above, the following steps are taken in data analysis of the  simulation results:

1. measure one-particle spectra for various sizes of lattice and moving frames,  and extract continuum limit particle mass, $m$,  by using relation \cite{Gatteringer:1993}: \mbox{$m(L) = m + \frac{c}{\sqrt{L}} e^{ - m L}$}.

2. measure two-particle spectra for various sizes of lattice  and moving frames, and extract the coupling strength, $V_0$,  from lattice data.  

3.   three-particle spectra are measured for various sizes of lattice  and moving frames as well,  three-particle spectra are thus compared with predicted three-particle spectra.  The predicted three-particle spectra, $E_{3b}^{(d)} (L)$, are given in terms of two independent particle momenta, $p_1$ and $p_2$, by 
\begin{equation}
 E^{(d)}_{3b}(L) = \sum_{i=1}^3 \cosh^{-1} \left ( \cosh m +1 - \cos p_i \right ),  \nonumber
\end{equation}
where $p_1$ and $p_2$ are the solutions of Eq.(\ref{3bexactsol}), and \mbox{$p_3 = P - p_1 -p_2$}.

  \begin{figure}
\begin{center}
\includegraphics[width=0.44\textwidth]{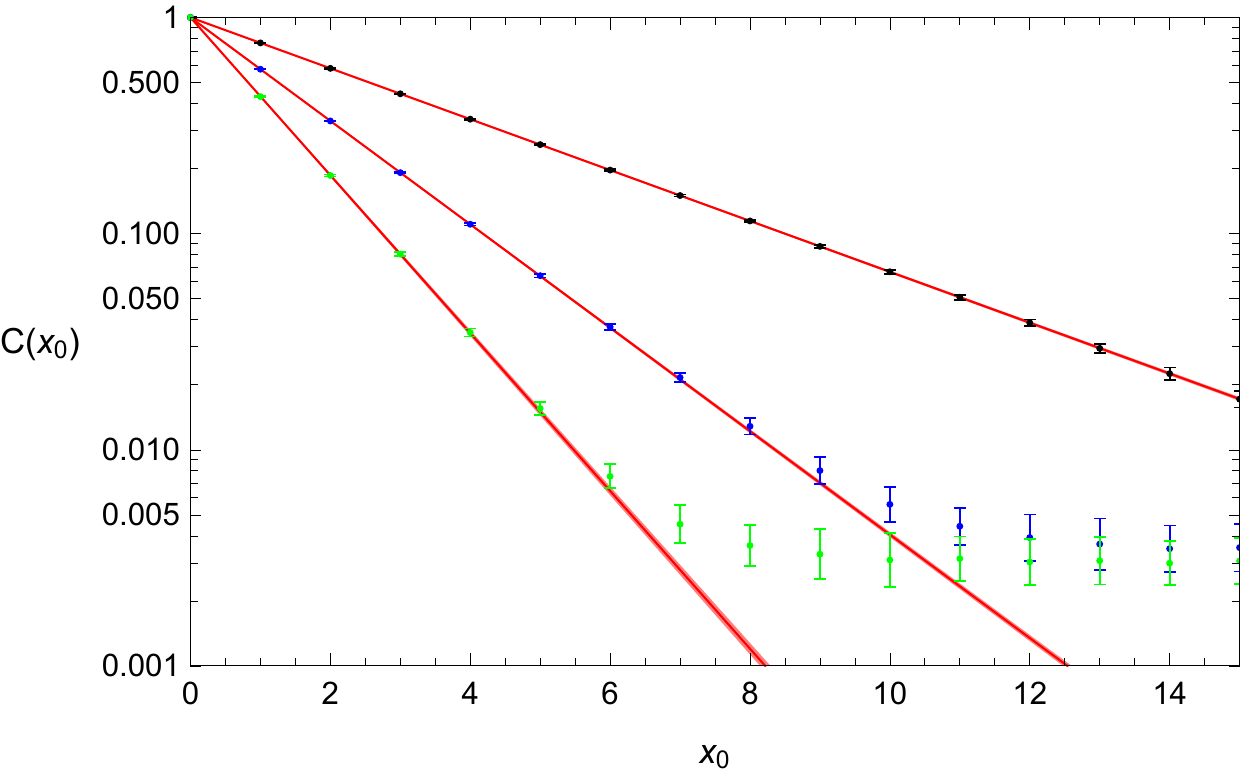}  
\caption{ Correlation functions  for one particle (black), two particles (blue) and three particles (green) at $L=40$ and $P=0$, and corresponding fitting curves (red band).  \label{corrplot}}
\end{center}
\end{figure}

  \begin{figure}
\begin{center} 
\includegraphics[width=0.44\textwidth]{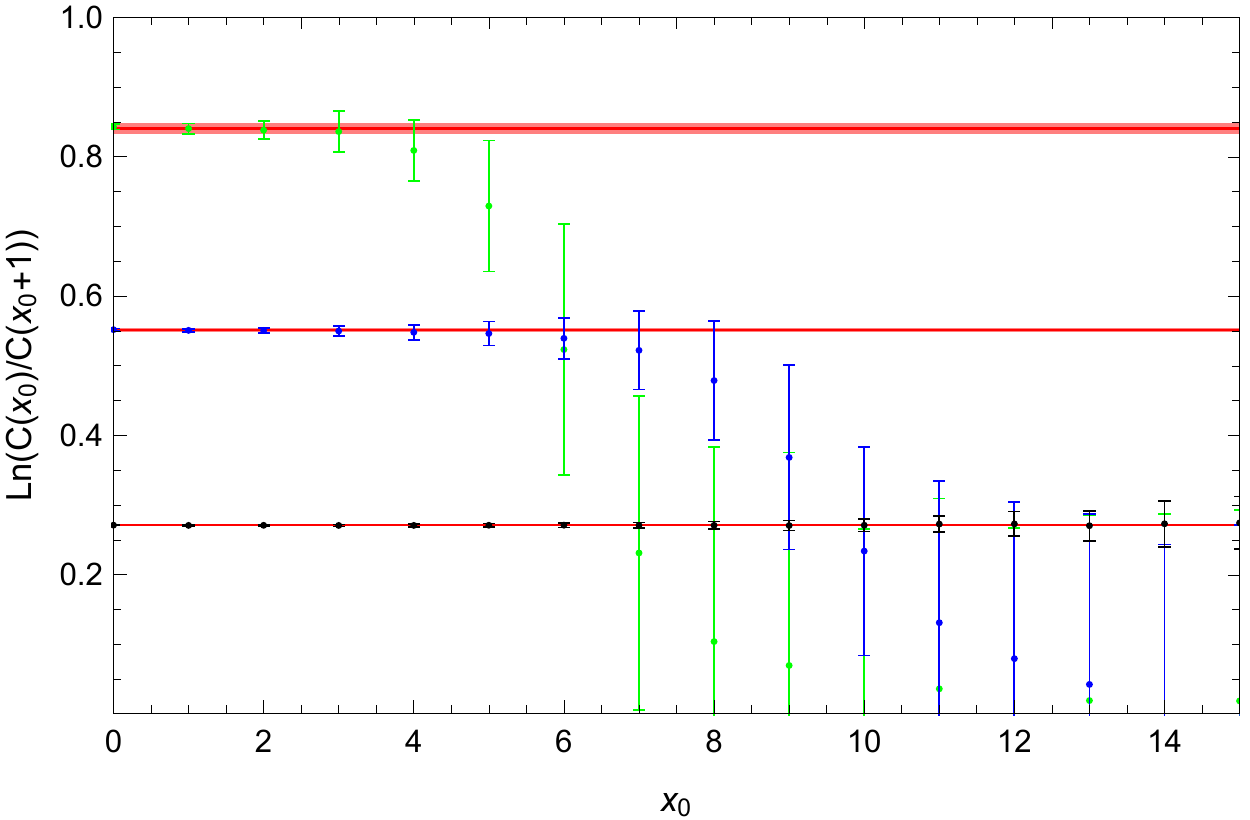}  
\caption{ Effective mass plots, $\ln \frac{C(x_{0})}{C(x_{0}+1)} $, for one particle (black), two particles (blue) and three particles (green) at $L=40$ and $P=0$, and corresponding fitting curves (red band). \label{effmassplot}}
\end{center}
\end{figure}

The particles spectra are extracted by fitting exponential   multi-particle correlation functions   as a function of time $x_{0}$:  $C(x_{0})\propto e^{- E x_{0}}$. See the example of one-, two- and three-particle correlation functions and effective mass, $\ln \frac{C(x_{0})}{C(x_{0}+1)}$,  in Fig.\ref{corrplot} and Fig.\ref{effmassplot}, respectively.  The construction of multi-particle operators and  correlation functions will be explained later  in Section \ref{spectra}.

\section{Particles spectra and data analysis}\label{spectra}

In this section, we present significant  results   for multi-particle scattering. Some details on multi-particle operator construction and data analysis are also given.

\subsection{One particle spectra}

\begin{figure}
\begin{center}
\includegraphics[width=0.44\textwidth]{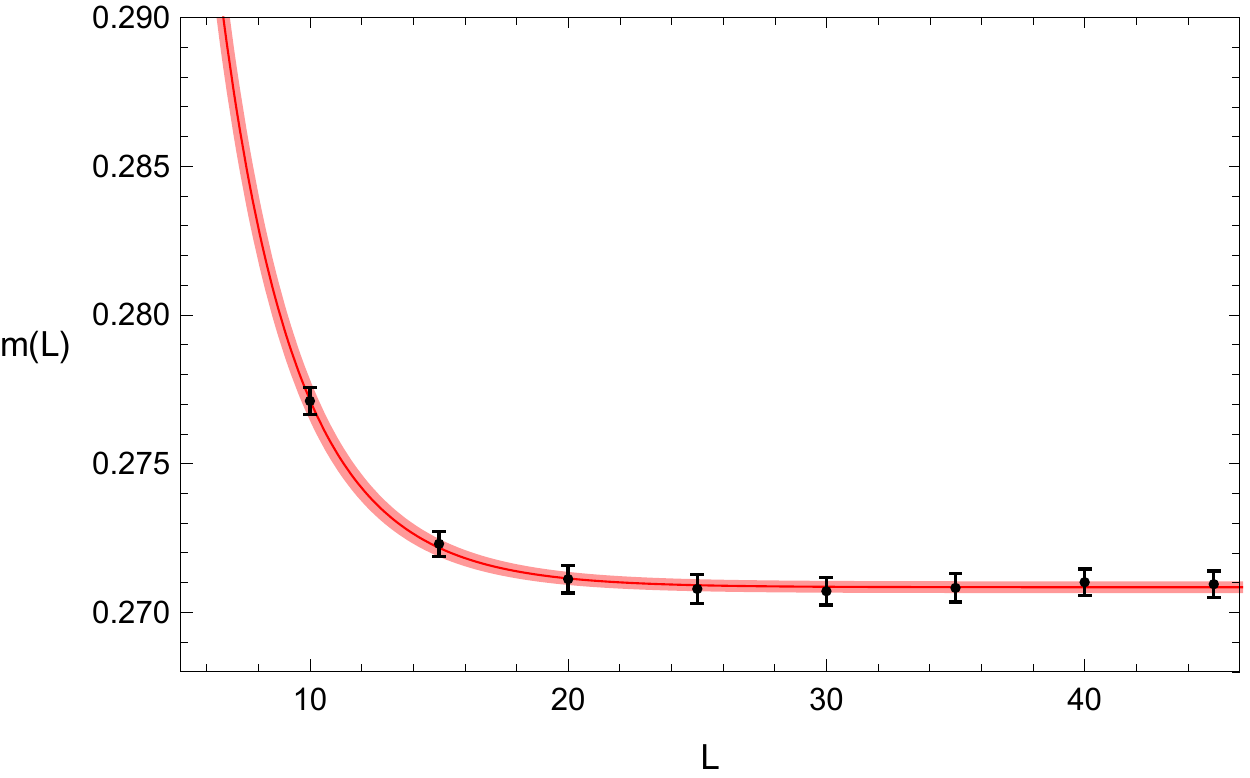}  
\caption{The single particle mass spectra $m (L) $ as function of lattice size $L$,  the single particle mass follows the relation: $m (L) = m + c /L^{1/2}e^{ - m L}$ (red band). \label{singlemass}}
\end{center}
\end{figure}

\begin{figure}
\begin{center}
\includegraphics[width=0.44\textwidth]{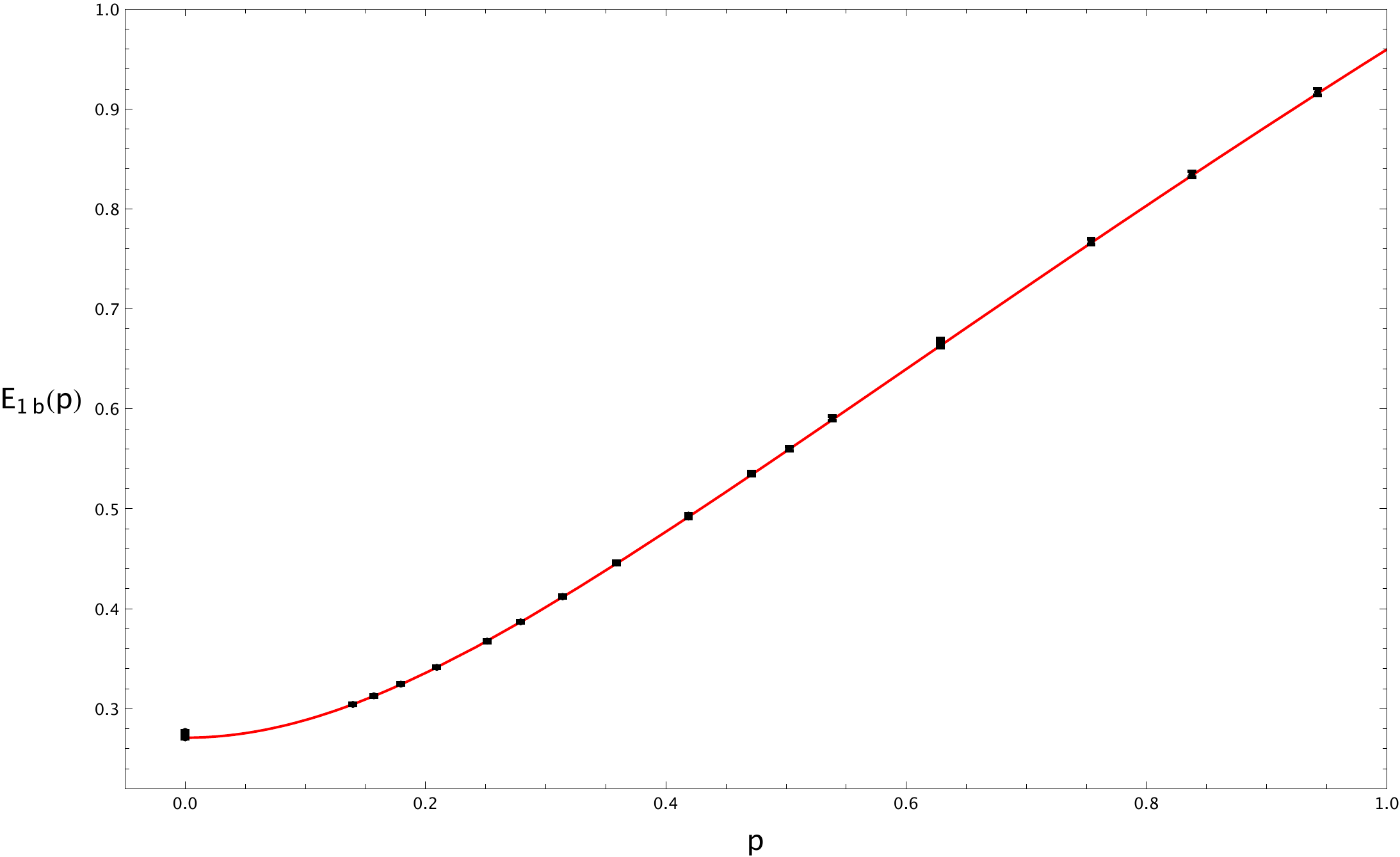}  
\caption{Plot of single particle spectra in various lattices    from $L=10$ up to $L=45$ vs.  lattice dispersion relation (red band),  \mbox{$E_{1b} (p) =  \cosh^{-1} \left ( \cosh m +1 - \cos p  \right )   $}, where  \mbox{$p=\frac{2\pi}{L}n , n \in \mathbb{Z}$}. \label{dispplot}}
\end{center}
\end{figure}

 The one-particle  spectra  are extracted from the exponential decay of the correlation functions
\begin{equation}
C_{ 1b, n}(x_{0}) = \langle   \widetilde{\phi}^*_{n}(x_{0}) \widetilde{\phi}_{n}(0) \rangle \propto e^{- E_{1b, n}  x_{0}},
\end{equation}
where  the one particle propagator, $\widetilde{\phi}_{n}(x_{0}) $, is defined by
\begin{equation}
\widetilde{\phi}_{n}(x_{0}) =\frac{1}{L} \sum_{x_{1}} \phi(x) e^{i x_{1} \frac{2\pi}{L} n },   \ \  n  \in \mathbb{Z}.
\end{equation}
Single particle energy $E_{1b,n} (L)$  is obtained for multiple lattice sizes, $L=10 $  up to $45$.   By fitting   single particle energies in multiple lattice sizes  with relation
\begin{equation}
m(L) = E_{1b,0}(L) = m + \frac{c}{\sqrt{L}} e^{- m L},
\end{equation}
where $c$ and $m$ are used as fitting parameters,   we thus find the mass of  single particle:  $m=0.2708 \pm 0.0002 $, see Fig.\ref{singlemass}. The excited single particle energy levels are used to check the energy-momentum dispersion relations in a finite lattice,
\begin{equation}
E_{1b,n}(L) =  \cosh^{-1} \left ( \cosh m +1 - \cos \frac{2\pi}{L} n  \right ).
\end{equation}
The comparison between lattice results and the lattice dispersion relation  is presented in Fig.\ref{dispplot}.  

\subsection{Two particles spectra}           
 
In   moving frames,  the matrix element of the two particle correlation function   read
\begin{equation}
C^{(d)}_{2b, (i, j)} (x_{0}) = \langle   O^{(d)*}_{2b, i}(x_{0}) O^{(d)}_{2b, j}(0)    \rangle ,
\end{equation}
where  $d\in \mathbb{Z}$ is related to center of mass momentum by $P= \frac{2\pi}{L} d$,   and  two-particle operators  are constructed by
\begin{align}
 O^{(d)}_{2b,n}(x_{0}) = \widetilde{\phi}_{n}(x_{0})   \widetilde{\phi}_{d-n}(x_{0}) .
\end{align}
Four two-particle operators are used in our simulation: $n=(0,1,2,3)$, so the size of matrix of two-particle correlation functions are   $4\times 4$, $3\times 3$ and $2\times 2$  for $d=0,1,2$.

\begin{figure}
\begin{center}
\includegraphics[width=0.49\textwidth]{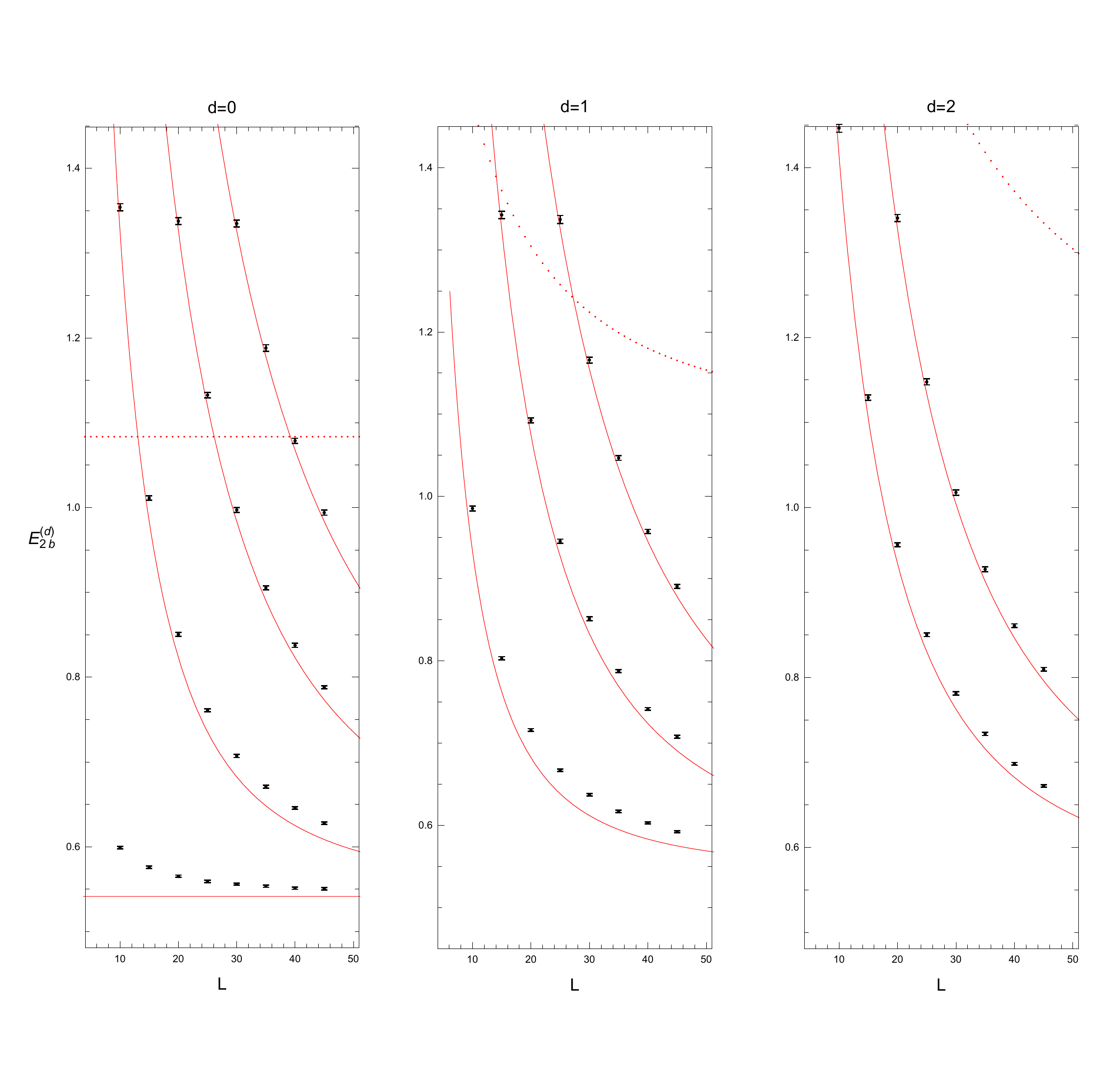}  
\caption{Plot of two particles spectra from various lattice size  from \mbox{$L=10$} up to \mbox{$L=45$} and \mbox{$d=0,1,2$} vs. free   two-particle energy levels (red curve):  \mbox{$E_{2b} (L)  = \sum_{i=1,2} \cosh^{-1} \left ( \cosh m +1 - \cos p_{i}  \right )   $}, where  \mbox{$p_{i}=\frac{2\pi}{L}n_{i} , n_{i} \in \mathbb{Z}$}. Dotted curves represent four-particle threshold. \label{E2bplot}}
\end{center}
\end{figure}

 The spectral decomposition of the correlation function matrices are usually given by  
\begin{equation}
C^{(d)}_{2b,( i,  j)} (x_{0}) = \sum_{n} v^{(d,n) *}_{2b, i}v^{(d,n) }_{2b, j} e^{- E_{2b, n}^{(d)}  x_{0}},
\end{equation}
where $v^{(d,n) }_{2b, i} =\langle n  |O^{(d)}_{2b, i}(0) |0\rangle $,  and $n$ labels the $n$-th energy eigenstate $E_{2b, n}^{(d)}$.  In order to extract excited energy states,  a generalized eigenvalue method  \cite{Luscher:1990ck}   is  proposed
\begin{equation}
C^{(d)}_{2b}(x_{0}) \xi_{2b,n} = \lambda^{(d)}_{2b, n} (x_{0}, \bar{x}_{0}) C^{(d)}_{2b}(\bar{x}_{0}) \xi_{2b, n} ,
\end{equation}
where    $\bar{x}_{0}$ is a small reference time.  Mixing of multi-particle states is protected by the conservation of charge quantum number in the complex $\phi^4$ model. Also,   diagonalized  correlation functions barely show the contamination of higher energy states in $ \lambda^{(d)}_{2b, n} (x_{0}, \bar{x}_{0})$, see Fig.\ref{corrplot} and Fig.\ref{effmassplot}. Therefore, $\bar{x}_{0}$  is set to zero and  a simple form of \mbox{$ \lambda^{(d)}_{2b, n} (x_{0},  0) = e^{-   E_{2b, n }^{(d)} x_{0} } $} is used in the data fitting for $x_0 \in [0,10]$. The two particle spectra for various lattice sizes and $d$ are presented in Fig.\ref{E2bplot}.

\begin{figure}
\begin{center}
\includegraphics[width=0.44\textwidth]{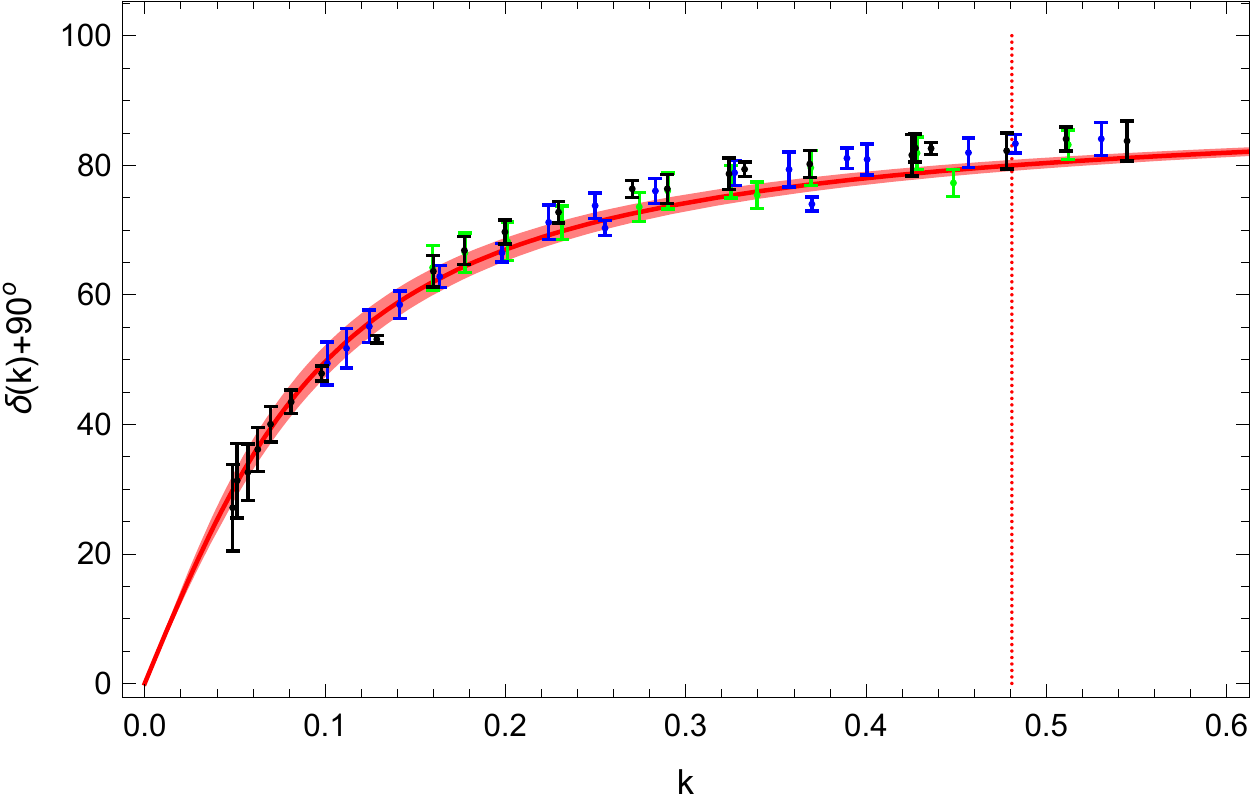}  
\caption{Two particles scattering phase shift $\delta_{lat}^{(d)} (k) $ from various lattices and several moving frames:  \mbox{$d=0$}(black), \mbox{$d=1$}(blue) and \mbox{$d=2$}(green),  vs. fitting result (red band) by using expression  $\delta (k) =\cot^{-1} \left ( - \frac{2 k}{mV_0} \right )$. The vertical line symbolize the four-particle threshold. \label{phaseshift}}
\end{center}
\end{figure}

The phase shift of two-body scattering is extracted from two-particle energy levels, $E_{2b, n }^{(d)} $, by  using relation:
\begin{align}
\delta_{lat}^{(d)} (k) = - \frac{k L}{2} - \frac{\pi }{2} d,
\end{align}
where the relative momentum of two particles, $k$, is given by the solutions of two-particle energy momentum dispersion relation
\begin{equation} 
E^{(d)}_{2b}(L) = \sum_{i=1,2} \cosh^{-1} \left ( \cosh m +1 - \cos p_i \right ), 
\end{equation}
where  $ p_1 = \frac{\pi}{L} d + k$ ,  $p_2 = \frac{\pi }{L} d - k $, see extracted phase shift in Fig.\ref{phaseshift}. The exact expression of phase shift, 
\begin{equation}
\delta (k) = \cot^{-1} \left ( - \frac{2 k}{mV_0} \right ),
\end{equation}
 is used to fit lattice results, $\delta_{lat}^{(d)} (k)$, and to fix the coupling strength, $V_0$. We thus find  $m V_0 = 0.170\pm 0.015$.

\subsection{Three particles spectra} 

For three-particle operators with \mbox{$d=0,1,2$},  four operators are used in present work:
\begin{equation}
 O^{(d)}_{3b, n}(x_{0}) = \widetilde{\phi}_{n}(x_{0})   \widetilde{\phi}_{-n}(x_{0}) \widetilde{\phi}_{d}(x_{0}) , \ \ n=0,1,2,
\end{equation}
and
\begin{align}
 O^{(d=0)}_{3b, 3}(x_{0}) & = \widetilde{\phi}_{1}(x_{0})   \widetilde{\phi}_{1}(x_{0}) \widetilde{\phi}_{-2}(x_{0}) ,  \nonumber \\
 O^{(d=1)}_{3b, 3}(x_{0}) & = \widetilde{\phi}_{0}(x_{0})   \widetilde{\phi}_{-1}(x_{0}) \widetilde{\phi}_{2}(x_{0}) ,  \nonumber \\
 O^{(d=2)}_{3b, 3}(x_{0}) & = \widetilde{\phi}_{0}(x_{0})   \widetilde{\phi}_{1}(x_{0}) \widetilde{\phi}_{1}(x_{0}) . 
\end{align}
Similar to the two-particle correlation function matrix, the   matrix element of the three particle correlation function   is given by
\begin{equation}
C^{(d)}_{3b, ( i, j)} (x_{0}) = \langle   O^{(d)*}_{3b, i}(x_{0}) O^{(d)}_{3b, j}(0)    \rangle .
\end{equation}
In the two-particle sector, the  generalized eigenvalue method  is also applied to extract three-body energy levels, 
 \begin{equation}
C^{(d)}_{3b}(x_{0}) \xi_{3b, n} = \lambda^{(d)}_{3b, n} (x_{0},  0) C^{(d)}_{3b}( 0 ) \xi_{3b, n} ,
\end{equation}
where \mbox{$ \lambda^{(d)}_{3b, n} (x_{0},  0) = e^{-   E_{3b, n }^{(d)} x_{0} } $}. An example of the three-particle correlation function,      $\lambda^{(d)}_{3b, n} (x_{0},  0) $, and effective mass, $\ln \left [ \lambda^{(d)}_{3b, n} (x_{0},  0) / \lambda^{(d)}_{3b, n} (x_{0}+1,  0) \right ] $ is given in Fig.\ref{corrplot} and Fig.\ref{effmassplot}.

\begin{figure}
\begin{center}
\includegraphics[width=0.49\textwidth]{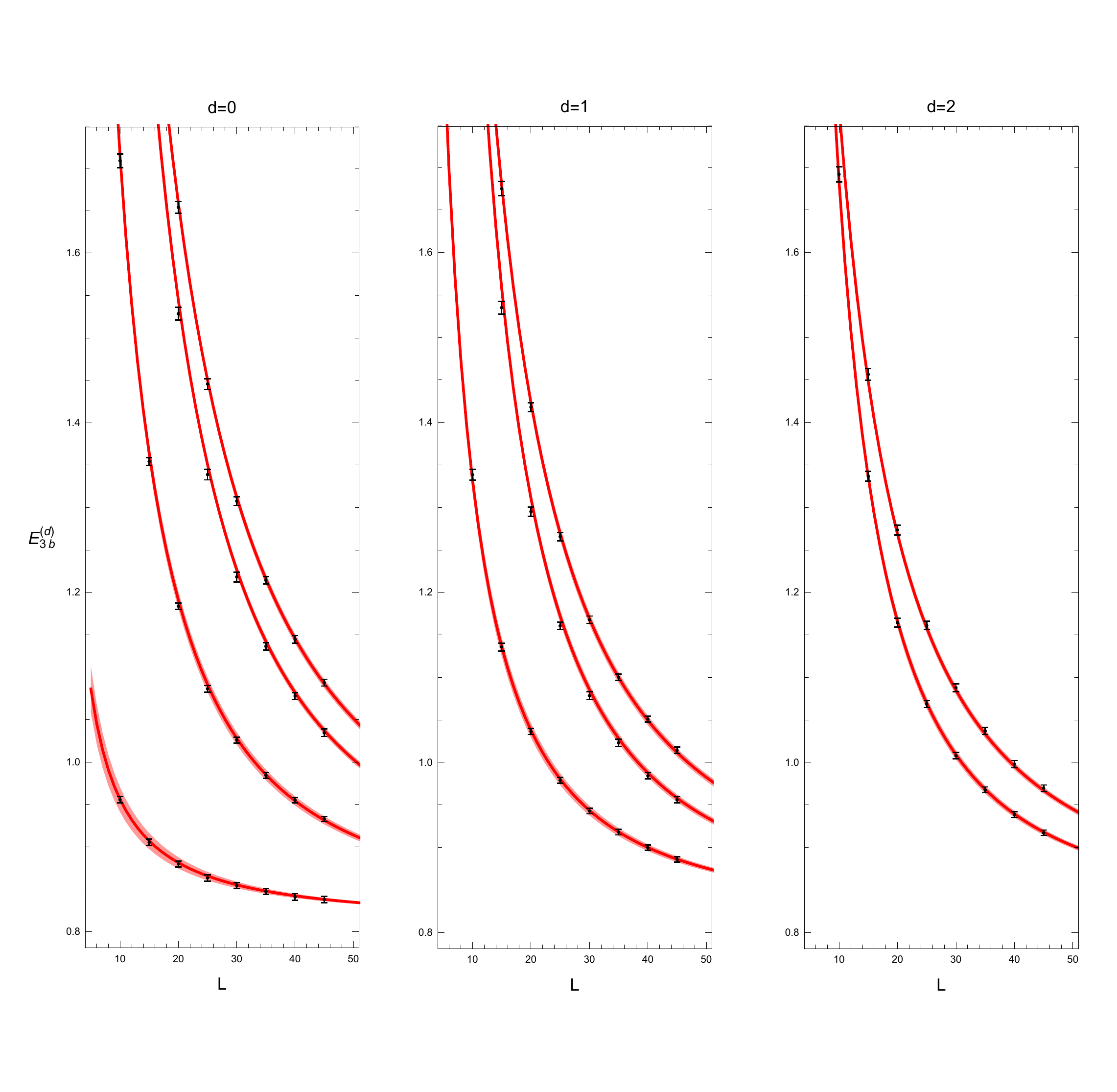}  
\caption{Three particles spectra from various lattice size  from \mbox{$L=10$} up to \mbox{$L=45$} and \mbox{$d=0,1,2$} vs. predicted three-particle energy levels (red band):  \mbox{$E_{3b} (L)  = \sum_{i=1}^{3} \cosh^{-1} \left ( \cosh m +1 - \cos p_{i}  \right )   $}, where  \mbox{$p_{3}=\frac{2\pi}{L} d -p_{1} - p_{2}$}, and the values of $p_{1}$ and $p_{2}$ are given by  solutions of  three-body quantization conditions   in Eq.(\ref{3bexactsol}). \label{E3bplot}}
\end{center}
\end{figure}

Given the values of particle mass, $m$, and coupling strength, $V_0$,  that we learned from discussion in previous sections,    three-particle spectra does not provide any new insight into the fundamental parameters of $\phi^4$ theory  due to the absence of three-body force.  However, in general, three-particle spectra are still  considered a  useful tool to explore and understand the dynamics of three-particle interaction.  In reality, it also provides opportunities to investigate the possibility of more fundamental parameters of  lattice QCD theory. Nevertheles, since the exact solutions are known, we only tend to demonstrate the consistence of predicted three-particle spectra compared to simulation results.  The predicted three-particle spectra are determined by three-body energy-momentum dispersion relations in terms of  two independent particle momenta, say $p_1$ and $p_2$,
\begin{equation}
 E^{(d)}_{3b}(L ) = \sum_{i=1}^3 \cosh^{-1} \left ( \cosh m +1 - \cos p_i \right ), 
\end{equation}
where  $p_3 = \frac{2\pi}{L} d  - p_1 -p_2$.  Two independent particle momenta, $p_1$ and $p_2$,  are the solutions of  three-body quantization conditions given in Eq.(\ref{3bexactsol}). The main results of three-particle spectra are presented in Fig.\ref{E3bplot}. As we can see,   the agreement  of predicted spectra against simulation results is quite remarkable.

\section{Summary }\label{summary}
In summary, the lattice simulation of multi-particle interaction is  studied by using a complex $\phi^{4}$ lattice model.  The simulation is performed in $1+1$ dimensional space and time for various sizes of lattices and multiple moving frames. The two dimensional $\phi^{4}$ model is exactly solvable and analytical expressions of multi-particle quantization conditions are known in finite volume \cite{Guo:2016fgl}. This feature makes it a perfect testbed for studying multi-particle interaction in a lattice. The typical $3-4$ numbers of multi-particle operators are used in our simulation, and a variational approach is implemented to extract excited state energy levels. Two parameters of $\phi^{4}$ theory, single particle mass and coupling strength, are extracted from single particle   and two particles spectra, respectively. Then, extracted $\phi^{4}$ theory parameters are applied to predict three-particle spectra by using analytical three-body quantization conditions compared with three-particle spectra from simulations. Predicted three-particle spectra and lattice results show quite remarkable agreement.

\section{ACKNOWLEDGMENTS}
 We   acknowledge support from the Department of Physics and Engineering, California State University, Bakersfield, CA.  We also thank Vladimir~Gasparian  and David~Gross for numerous fruitful discussions.   This research was supported in part by the National Science Foundation under Grant No. NSF PHY-1748958.

\end{document}